\documentclass[12 pt]{article}
\usepackage{ucs} 
\usepackage[utf8x]{inputenc} 
\usepackage[russian,english]{babel}  

\usepackage{amsmath, amsfonts, amssymb}
\usepackage{graphicx}
\usepackage{indentfirst}
\usepackage{threeparttable}
\usepackage{url}

\def\la{\;
\raise0.3ex\hbox{$<$\kern-0.75em\raise-1.1ex\hbox{$\sim$}}\; }
\def\ga{\;
\raise0.3ex\hbox{$>$\kern-0.75em\raise-1.1ex\hbox{$\sim$}}\; }

\newcommand{\kms}{km~s$^{-1}$}

\newcommand{\dmm}{$\Delta\mu/\mu$}
\newcommand{\chhhoh}{CH$_3$OH}

\begin{document}

\title{\LARGE \bf 
Indications of electron-to-proton mass
ratio variations in the Galaxy. III. 0.6 mm
methanol lines toward Sgr\,B2(N) and Orion-KL.
}

\maketitle

\begin{center}
J. S. Vorotyntseva\footnote{e-mail: j.s.vorotyntseva@mail.ioffe.ru}, 
S. A. Levshakov \\
Ioffe Institute, St. Petersburg, 194021 Russia
\end{center}

\bigskip

\begin{abstract} 
In this paper, we show that  methanol (CH$_3$OH)  
torsional-rotational transitions, which have increased 
sensitivity to small variations 
of $\mu$~-- the electron-to-proton mass ratio, are shifted relative 
to less sensitive transitions in the spectrum of the Sgr\,B2(N) molecular 
cloud located at the Galactic center.
At the same time, an identical set of methanol lines in the spectrum of 
the Orion-KL molecular cloud, located far from the center, shows no shifts.
Interpreting this behavior of molecular frequencies in terms of
$\Delta\mu/\mu = (\mu_{\rm obs} - \mu_{\rm lab})/\mu_{\rm lab}$
leads to the following weighted mean values:
$\langle \Delta\mu/\mu \rangle = (-3.4\pm0.4)\times10^{-7}$ for Sgr\,B2 
and
$\langle \Delta\mu/\mu \rangle = (-1.1\pm0.8)\times10^{-7}$ for Orion-KL
(indicated are the total $\pm 1\sigma$ errors  
of the  weighted means $\langle \Delta\mu/\mu \rangle$
including both
statistical and systematic uncertainties).
A possible correlation between \dmm\ values measured in various
molecular clouds of the Galactic disk
and the distribution of dark matter along the Galactic radius is discussed,
which may suggest a hypothetical  modulation of the Higgs scalar field
by dark matter, resulting in a change of $\mu$.
\end{abstract}

\section{Introduction}
\label{Sec1}

Complex organic molecules such as methanol (\chhhoh),
ubiquitous in
molecular clouds that fill the Galactic disk from
the circumnuclear regions with galactocentric distances of 
$R_{\scriptscriptstyle\rm GC} \la
100$~pc, like Sagittarius (Sgr) B2 \cite{Nei14, Mil18}, to the periphery
($R_{\scriptscriptstyle\rm GC} \la 23$~kpc \cite{Ber21, Bra23}) 
are convenient
probes not only of the physical conditions in star-forming regions,
but also a unique tool in the search for new physics
(see, e.g., the review \cite{KL13}).
The special role of methanol in these tasks is explained by three factors:
($i$) the frequencies of torsional-rotational transitions 
in \chhhoh\ exhibit 
high sensitivity to small variations in the parameter
$\mu = m_{\rm e}/m_{\rm p}$~-- the electron-to-proton mass 
ratio \cite{LKR, Jan};
($ii$) the methanol line spectrum fills a broad frequency 
range from GHz to THz, 
accessible to observations with modern ground-based 
and orbital telescopes; 
($iii$) the methanol lines are sufficiently intense to provide
high-quality spectra with a high signal-to-noise ratio.

Let us now explain what is meant by the term ``New Physics''.
According to Einstein's Equivalence Principle (EEP)
and its fundamental component, the Local Position Invariance (LPI), 
the results of non-gravitational experiments should be independent 
of spatial and temporal coordinates.
The detection of deviations from the LPI in laboratory 
experiments with atomic and nuclear clocks and/or in 
astrophysical spectral observations would mean the 
manifestation of new physical laws that go beyond the Standard Model of particle physics.
A violation of the EEP is assumed in various chameleon scenarios
of the screening of the fifth force,
in multidimensional models, in models of the dark sector 
(dark matter, DM, and dark energy, DE)
 and in a number of other theories, considered, for example, in the review
\cite{Uz25}.

In practical tasks of searching for New physics,
spectral lines of various chemical elements are widely used,
since the structure of the energy levels of atoms and molecules
depends on the values of fundamental physical
constants, which, if the LPI is violated, would change their values,
thereby causing shifts in the corresponding frequencies.
In particular, small changes in the constant $\mu$
would lead to significant shifts in the frequencies of tunnel
transitions in molecules such as NH$_3$, \chhhoh\, etc. \cite{KL13}, 
\cite{VKL24}-\cite{VLK26}.

The masses of elementary particles (an electron and three quarks), 
on which
the ratio $m_{\rm e}/m_{\rm p}$ depends,
are in turn modulated by the Higgs scalar field.
However, the proton mass  (938 MeV/$c^2$) is determined 
mainly by the binding energy of three quarks, 
the total mass of which ($\sim 10$ MeV/$c^2$)
is approximately 1\% of the total mass of the proton.
Therefore, it can be assumed that, first of all,
the variations of $\mu$ should have been associated with 
a change in the electron mass,
which is given as the product of the Yukawa coupling constant 
for the electron $\lambda_{\rm e}\approx 2.9\times10^{-6}$
and the mass of the Higgs boson $v$ ($\sim 246$ GeV):
$m_{\rm e} = \lambda_{\rm e}v/\sqrt{2}$ \cite{Ca17}.

The cosmological evolution of the Higgs boson mass $v$
in the expanding Universe, and hence the 
dependence of $\mu$ on cosmological time, 
was considered, for example, in \cite{Ca17}.
The question we will attempt to answer in this paper is 
the local behavior of $\mu$, namely, whether there is any
correlation between the local dark matter density 
$\rho_{\scriptscriptstyle\rm DM}$
and the measured value of $\Delta\mu/\mu =
(\mu_{\rm obs} - \mu_{\rm lab})/\mu_{\rm lab}$
in the Galactic disk.

In the Galaxy, as is well known,
with increasing $R_{\scriptscriptstyle\rm GC}$, 
the ratio $\rho_{\rm br}/\rho_{\scriptscriptstyle\rm DM}$ 
changes significantly:
the density of the baryon component $\rho_{\rm br}$ sharply
decreases with increasing
$R_{\scriptscriptstyle\rm GC}$, 
while $\rho_{\scriptscriptstyle\rm DM}$ 
increases \cite{Sof}-\cite{Tot}.
Both densities equalize at approximately two effective radii,
($R_{\rm eff}$)\footnote{The effective radius ($R_{\rm eff}$) 
is defined as the length
of the semimajor axis of the elliptical isophote, which includes 
half the light
of the galaxy's stellar component. For the Galaxy,
$R_{\rm eff} \approx 4.5$ kpc,
and the galactocentric distance of the solar system
$R_\odot \approx 1.8R_{\rm eff}$ \cite{Li15, Zh23, Isi25}.},
after which dark matter becomes the dominant component of the
Galactic disk and halo.

Limits on the variations of $\mu$ far from the Galactic 
center ($R_{\scriptscriptstyle\rm GC} \ga R_{\rm eff}$)
were previously established using different methods.
The tightest limits were obtained at the level of
$\Delta\mu/\mu < (1-4)\times10^{-8}$ \cite{VLK24}-\cite{VLK26},
\cite{E11}-\cite{L22}.
The Galactic center remained unexplored until recently.

First attempts to estimate $\Delta\mu/\mu$ using three methanol transitions at frequencies $\sim$ 540 GHz (observations of the {\it Herschel} space telescope\footnote{{\it Herschel} is an ESA space observatory,
equipped with instruments developed by leading European centers
with significant participation from NASA.},
and also eleven methanol lines from the 3 mm atmospheric window (observations with the 30-m IRAM telescope\footnote{The Institute for
Radio Astronomy in the Millimeter Range (IRAM) is an international
research institute and Europe's leading center for radio astronomy.})
showed a possible decrease in $\mu$ in the molecular complex Sgr\,B2(N): 
$\Delta\mu/\mu = (-4.2 \pm 0.7)\times10^{-7}$ \cite{VL2025}
and $\Delta\mu/\mu = (-2.1 \pm 0.6)\times10^{-7}$ \cite{VLH},
respectively.

In this paper, we continue our studies in the high-frequency range 
(490-640 GHz) based on observations by the {\it Herschel} 
space telescope in two directions: toward the Galactic Center, the Sgr B2(N) molecular complex, and, for control, toward an object localized far from the Galactic center at a distance of
 $R_{\scriptscriptstyle\rm GC} \sim 9$ kpc~-- the Orion-KL 
 molecular complex.
The new analysis was performed on an expanded sample of methanol lines
compared to \cite{VL2025}: ten new lines, selected from each of 
the {\it Herschel} spectra of these two objects, were added to the three previous lines.

\section{Parameters of {\it Herschel} observations} 
\label{Sec2}

In this work, the archival spectra
of the {\it Herschel} space observatory, obtained
in the direction of two molecular complexes ~-- Sgr\,B2(N) and Orion-KL, are used.
Both objects belong to
active star formation zones with masses and linear sizes
$M_{\scriptscriptstyle\rm Sgr\,B2(N)} \sim 2\times10^3 M_\odot$,
$D_{\scriptscriptstyle\rm Sgr\,B2(N)} \sim 0.2$ pc \cite{Sch19} and
$M_{\scriptscriptstyle\rm Ori-KL} \sim 2\times10^2 M_\odot$,
$D_{\scriptscriptstyle\rm Ori-KL} \sim 0.02$ pc \cite{Pag17}.

A description of the equipment used to obtain the spectra, as well as information about the primary processing of the spectra, is given in our work \cite{VL2025}.
Here we report only the main technical
characteristics listed in Table~\ref{Tab1}:
observed frequency ranges, corresponding spectral bands, telescope guidance coordinates
(right ascension and declination), dates of observations, aperture sizes
 $\theta_{\scriptscriptstyle\rm HPBW}$
and spectral resolution $\Delta_{\rm ch}$.

\begin{table*}[h!]
\centering
\caption{Characteristics of  observations
and coordinates of the objects \cite{Nei14, Cro15}.
The second column shows the frequency ranges $\Delta f$ in which the methanol lines \chhhoh\ were selected.
The corresponding observed frequency bands have
the following values:
 $479.5-561.5$ GHz ({\it band 1a}),
$554.5-635.5$ GHz ({\it band 1b}) and $626.0-726.0$ GHz ({\it band 2a}).
Telescope aperture $\theta_{\scriptscriptstyle\rm HPBW}$
(Half Power Beam Width) in the center of the band and spectral resolution
$\Delta_{\rm ch}$ (channel width) are given in the fifth and sixth columns.
For all three spectral ranges, the width of the intermediate 
frequency band (IF) is equal to 4 GHz \cite{Nei14, HIFI}.
}
\label{Tab1}
\begin{tabular}{l c c c c r@.l}
\hline
\multicolumn{1}{c}{Target} & Frequency & Coordinates& Observation &
$\theta_{\scriptscriptstyle\rm HPBW}$& \multicolumn{2}{c}{$\Delta_{\rm ch}$}\\
&Range, $\Delta f$ &(J2000)& Date & ($^{\prime\prime}$) & 
\multicolumn{2}{c}{(\kms)} \\
\hline\\[-10pt]
{\bf Sgr\,B2(N)} &
493--496 GHz&$17^{\rm h}47^{\rm m}20.\!^{\rm s}06$
&22.09.2010&40.4&0&3\\
&541--544 GHz   &$-28^{\circ}22^{\prime}18.\!^{\prime\prime}33$&&
 \multicolumn{2}{c}{}\\
&({\it band 1a})&&&\multicolumn{2}{c}{}\\[5pt]

&589--591 GHz&$17^{\rm h}47^{\rm m}20.\!^{\rm s}04$
&12.10.2010&35.3&0&25\\
&({\it band 1b})&$-28^{\circ}22^{\prime}18.\!^{\prime\prime}29$&&
 \multicolumn{2}{c}{ }\\[5pt]

&637--639 GHz&$17^{\rm h}47^{\rm m}19.\!^s{\rm}55$
&15.09.2010&31.1&0&23\\
&({\it band 2a})&$-28^{\circ}22^{\prime}18.\!^{\prime\prime}33$&&
 \multicolumn{2}{c}{}\\[10pt]

{\bf Orion-KL} &
493--496 GHz&$05^{\rm h}35^{\rm m}14.\!^{\rm s}36$ 
& 01.03.2010 & 40.4 & 0&3 \\
&541--544 GHz   &$-05^{\circ}22^{\prime}33.\!^{\prime\prime}63$&&
 \multicolumn{2}{c}{}\\
&({\it band 1a})&&&\multicolumn{2}{c}{}\\[5pt]

&589--591 GHz&$05^{\rm h}35^{\rm m}14.\!^{\rm s}35$
&02.03.2010&35.3 &0&25\\
&({\it band 1b})&$-05^{\circ}22^{\prime}33.\!^{\prime\prime}06$&&
 \multicolumn{2}{c}{}\\[5pt]

&637--639 GHz&$05^{\rm h}35^{\rm m}14.\!^{\rm s}32$
&12.04.2010&31.1 &0&23\\
&({\it band 2a})&$-05^{\circ}22^{\prime}32.\!^{\prime\prime}91$&&
 \multicolumn{2}{c}{}\\
\hline
\end{tabular}
\end{table*}

Among the spectra corresponding to these ranges, there are a number of methanol  (CH$_3$OH) lines  observed in both Sgr\,B2(N) and Orion-KL, which are most suitable for estimates of $\Delta\mu/\mu$.
The selection criteria for such lines are as follows:
 (1) a sufficient signal-to-noise ratio of $SNR\geq 40$; 
 (2) lines falling into the same IF band (4 GHz);
(3) close values of the energies of the upper levels  $E_{u}$; 
(4) different sensitivity coefficients $Q$ (see Table~\ref{Tab2})
to small variations of $\mu$ ($|\Delta Q|> 0$); 
(5) known laboratory frequency $f_{\rm lab}$.
This selection and distribution of lines across
sub-ranges is chosen to avoid possible systematic errors associated with the calibration of the frequency and temperature scales.
First of all, the calibration error of the frequency scale is eliminated, 
which can be significant for different IF bands and different 
spectral ranges.
Another source of instrumental errors is different observation dates and different aperture sizes.
Also, the selected lines do not fall into the marginal regions of the spectral band in order to avoid the known 
the ``accordion effect"~-- changes of channel widths at the edges of the spectral range.
The median error of calibration of the frequency scale
within the IF band does not exceed 50 kHz \cite{Av}.
The temperature scale ($T_{\scriptscriptstyle\rm A}$) was calibrated
with an accuracy of 1.6-4\% in band 1 and 1.3-3\% in band 2 (see Table~5.6 in \cite{HIFI}).

The parameters of the methanol transitions selected in this way are summarized in Table~\ref{Tab2}:
the first column shows the line number, the second~-- quantum numbers~--
the total angular momentum $J$ and
its projection onto the axis of symmetry of the molecule $K$ for the upper $u$ and lower $\ell$ levels, the third column lists the energies of the upper levels $E_u$, taken from \cite{Xu1997}, the fourth and fifth columns display the laboratory frequencies $f_{\rm lab}$\cite{Xu1997} and the calculated frequencies $f_{\rm cal}$\cite{Xu2008}, respectively. 
The  values of the sensitivity coefficients $Q$\cite{Jan} are given in the
last column.

\begin{table*}[h!]
\centering
\caption{Parameters of the selected methanol lines CH$_3$OH.
The measured laboratory $f_{\rm lab}$ and calculated $f_{\rm cal}$ 
line frequencies (in MHz)
are taken, respectively, from \cite{Xu1997} and \cite{Xu2008}.
The energies of the upper torsion-rotational levels $E_u$ are indicated in
the third column.
The values of the sensitivity coefficients $Q$ are calculated in \cite{Jan},
where $K_\mu = -Q$.
The uncertainties of 1$\sigma$
in the last digits are given in parentheses.
}
\label{Tab2}
\begin{tabular}{c l r@.l c c r@.l }
\hline
\# &\multicolumn{1}{c}{Transition} & \multicolumn{2}{c}{$E_u$} & 
Measured & Calculated &  \multicolumn{2}{c}{$Q$} \\
& $J_{K_u} \to J_{K_\ell}$ & \multicolumn{2}{c}{(K)} & frequencies, $f_{\rm lab}$ &
frequencies, $f_{\rm cal}$ \\
\hline\\[-10pt] 

1&{$5_{3} \to 4_{2}A^+$}& 84&6 & 493699.095(50) & 493699.112(15) &--0&1\\
2&{$5_{3} \to 4_{2}A^-$}&  84&6 & 493733.672(50) & 493733.687(15) &--0&1\\
3&{$7_{2} \to 7_{1}A^{-+}$}& 102&7 & 494481.683(50) & 494481.555(12) &1&5\\ 
4&{$7_{0} \to 6_{-1}E$}& 78&0 & 495173.104(50)&495173.105(12) &1&8\\[5pt]

5&{$6_{3} \to 5_{2}A^+$}& 98&5 & 542000.981(50)&542000.954(14) &0&0\\
6&{$6_{3} \to 5_{2}A^-$}&  98&5 & 542081.936(50)& 542081.952(14) &0&0\\
7&{$8_{0} \to 7_{-1}E$}&  96&6 & 543076.194(50)& 543076.175(12) &1&7\\ [5pt]

8 &{$7_{3} \to 6_{2}A^+$}& 115&0 & 590277.688(50)& 590277.712(14) &0&1\\
9&{$7_{3} \to 6_{2}A^-$}& 114&7 & 590440.291(50)&  590440.430(14) &0&1\\
10&{$9_{0} \to 8_{-1}E$}&117&4  & 590790.957(50)&  590790.939(12)&1&6\\[5pt]

11&{$10_{0} \to 9_{-1}E$}& 140&5& 638279.564(50)&638279.651(12) &1&6\\ 
12&{$8_{3} \to 7_{2}A^+$}& 133&3 & 638523.486(50)&638523.509(13) &0&1\\
13&{$8_{3} \to 7_{2}A^-$}& 133&3 & 638817.830(50)&638817.824(13) &0&1\\ 
\hline
\end{tabular}
\end{table*}

\section{Spectra processing and calculated line parameters}
\label{Sec3}

The processing of the spectra consisted of describing the shape of the profile of the observed line using an envelope curve and determining its
center~-- the point at which the envelope reaches its maximum.
For this purpose, we used a set of Gaussians with the minimum number of components, which provided the value
$\chi^2_\nu\la 1$, where $\nu$ is the number of degrees of freedom.

\begin{figure}
\vspace{-4.0cm}
\centering
\includegraphics[width=1.0\textwidth]{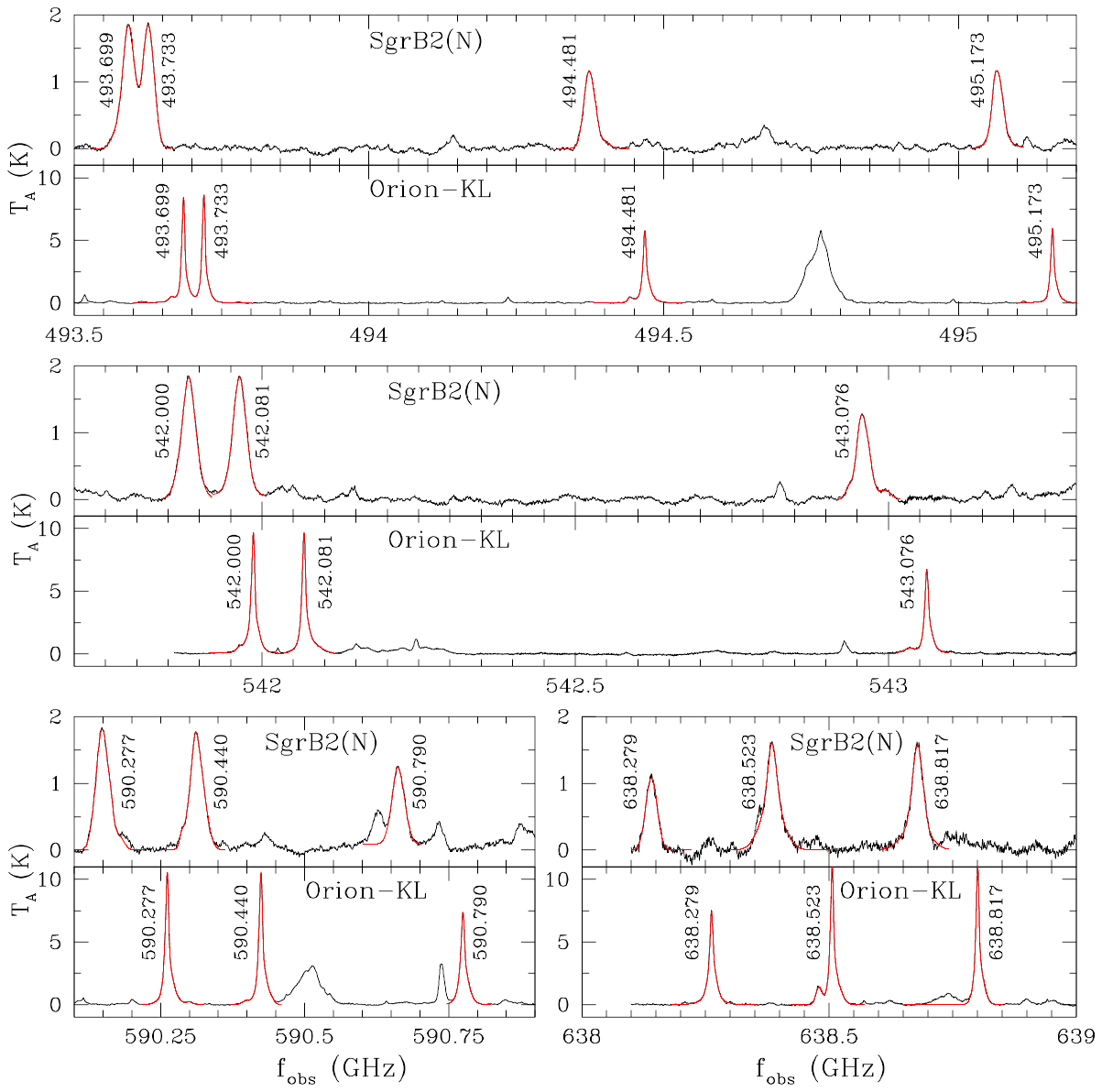}
\vspace{-3.5cm}
\caption{\small 
The methanol (\chhhoh) lines toward two sources~-- Sgr\,B2(N) and 
Orion-KL, obtained
at the space observatory {\it Herschel}.
The original spectra are shown in black
and the model spectra~-- in red.
The numbers next to the lines indicate laboratory frequencies 
$f_{\rm lab}$ in GHz (see Table~\ref{Tab2}).
}
\label{Fig1}
\end{figure}

The envelope curves calculated in this way are shown in Fig.~\ref{Fig1} 
in red and the original spectra~-- in black. An object and the laboratory frequencies of the observed lines are signed for each spectrum.

The calculated parameters of the methanol lines from the Sgr\,B2(N) spectrum are listed in Table~\ref{Tab3}: serial number of the line 
(according to Table~\ref{Tab2})~-- in the first column,
the observed frequency (measured center of the line) $f_{\rm obs}$~-- 
in the second column with the designation of the type of molecule~-- 
A or E,
the third and fourth columns show the widths ($FWHM$)
of the  lines and the observed peak temperatures of the main 
antenna beam 
$T^{\scriptscriptstyle\rm peak}_{\scriptscriptstyle\rm A}$
respectively,
the radial velocities 
 $V^{\scriptscriptstyle(k)}_{\scriptscriptstyle\rm LSR}$
are given in the fifth and sixth columns which are calculated in accord with the radio astronomical convention: 
\begin{equation}
V^{\scriptscriptstyle (1)}_{\scriptscriptstyle\rm LSR} = c \left( 1 - {f_{\rm obs}}/{f_{\rm lab}} \right),
\,\,\,
V^{\scriptscriptstyle (2)}_{\scriptscriptstyle\rm LSR} = c \left( 1 - {f_{\rm obs}}/{f_{\rm cal}} \right).
\label{E3}
\end{equation}

The line width errors are $\sim 5$\%, taking into account the contribution from the uncertainty of the flux calibration  and the contribution from the measurement errors of the line widths from the observed spectra.
The uncertainties of peak temperatures include 
the statistical component~-- the average local
noise amplitude (rms) near the methanol line under consideration~-- and the systematic~-- majority
error of the temperature scale calibration (see Sect.~\ref{Sec2}).
Both contributions were added up quadratically.
The radial velocity uncertainties
were calculated from the statistical error of determining 
the center of the line
$\sigma_{f_{\rm obs}}$, reference
frequency errors $\sigma_{f_{\rm lab}}$, as well as from
the median calibration error of the  frequency scale, 50 kHz \cite{Av}.
The statistical uncertainties $\sigma_{f_{\rm obs}}$ were calculated  
 using the $\Delta\chi^2$ method \cite{Press}, taking into account the correlation of noise in the spectra.
The correction factor $\gamma \sim 1.15$ 
(equation (5) in \cite{VLH}) slightly adjusted the noise amplitude
$\sigma^{\rm cor}_{\rm rms} = \gamma \sigma_{\rm rms}$.
This method of calculating the statistical error is justified, since the calculated values $\sigma_{f_{\rm obs}}$ match with good accuracy the expected values, which are expressed in terms of channel width 
$\Delta_{\rm ch}$, the full line width $FWHM$ and 
the signal-to-noise ratio $SNR$ 
(see formula (8) and Table 3 in \cite{VL2025}).

The exception is lines 1 and 2 from the Sgr\,B2(N) spectrum, which turned
out to be partially resolved (see Fig.~\ref{Fig1}).
The wings of these lines
influence the definition of the envelope centers,
shifting them either to the left (line 1) or
to the right (line 2) by almost the channel width relative to the mean value
$V_{\scriptscriptstyle\rm LSR} = 65.138$ km~s$^{-1}$, despite the fact that both lines have the same sensitivity coefficients $Q = -0.1$.
Similar lines in the Orion-KL spectrum, where there is no wing overlap, show the same radial velocities (see Table~\ref{Tab4}).
Therefore, we increased the errors of lines 1 and 2
in the Sgr\,B2(N) spectrum to the channel width 
$\Delta_{\rm ch} = 0.3$ km~s$^{-1}$.
In all other cases, the resulting radial
velocity error is calculated as the quadratic sum of the errors
 $\sigma_{f_{\rm obs}}$, $\sigma_{f_{\rm lab}}$
and systematic error
$\sigma_{f_{\rm sys}} = 50$ kHz, which were discussed above.

\begin{table*}[h!]
\centering
\caption{Parameters of the observed methanol lines in the 
Sgr\,B2(N) spectrum~--
center frequency $f_{\rm obs}$, full line width at half maximum 
$FWHM$ and the peak intensity 
in the center of the line 
$T^{\scriptscriptstyle\rm peak}_{\scriptscriptstyle\rm A}$.
The lines are numbered according to Table~\ref{Tab2}.
$V^{\scriptscriptstyle (1)}_{\scriptscriptstyle\rm LSR}$ and
$V^{\scriptscriptstyle(2)}_{\scriptscriptstyle\rm LSR}$~-- radial velocities
calculated using measured laboratory frequencies $f_{\rm lab}$ and
predicted in theoretical calculations $f_{\rm cal}$,
respectively (see Table~\ref{Tab2}).
The $1\sigma$ errors in the last digits are shown in parentheses.  
}
\label{Tab3}
\begin{tabular}{c r@.l  c c r@.l r@.l}
\hline\\[-12pt]
\# & \multicolumn{2}{c}{$f_{\rm obs}$} & $FWHM$ & 
$T^{\scriptscriptstyle\rm peak}_{\scriptscriptstyle\rm A}$ &
 \multicolumn{2}{c}{$V^{\scriptscriptstyle (1)}_{\scriptscriptstyle\rm LSR}$} &
\multicolumn{2}{c}{$V^{\scriptscriptstyle (2)}_{\scriptscriptstyle\rm LSR}$} \\
 & \multicolumn{2}{c}{(MHz)} & (\kms) & (K) & 
  \multicolumn{2}{c}{(\kms)} & \multicolumn{2}{c}{(\kms)} \\
\hline\\[-10pt]

1&493592&41(50) A& 15 & 1.85(8) & 64&78(30) & 64&79(30)  \\
2&493625&81(50) A& 14 & 1.85(8) & 65&49(30) & 65&50(30) \\
3&494373&85(10) A& 14 & 1.16(5) & 65&38(7) & 65&30(7) \\
4&495064&98(7) E& 13  & 1.16(5) & 65&46(6) &  65&47(5) \\[5pt]

5&541883&19(6) A& 15 & 1.83(8) & 65&15(5) & 65&14(4) \\
6&541964&24(6) A& 14 & 1.82(8) &  65&09(5) & 65&10(4) \\
7&542957&84(8) E& 13 & 1.27(6) & 65&34(6)  &  65&32(5) \\[5pt]

8&590148&89(6) A& 14  & 1.81(8) & 65&41(5) &  65&43(4) \\
9&590311&39(6) A& 13  & 1.76(8) & 65&45(5) &  65&52(4)\\
10&590661&67(9) E& 13 & 1.24(6) & 65&61(6) & 65&60(5) \\[5pt]

11&638140&50(20) E& 13 & 1.08(7) & 65&32(10) &  65&36(10) \\
12&638384&62(18) A& 14  & 1.60(8) & 65&20(9) &  65&21(9) \\
13&638679&10(17) A& 13 & 1.57(8) & 65&11(9) & 65&10(8) \\
\hline
\end{tabular}
\end{table*}

\begin{table*}[h!]
\centering
\caption{Same as Table~\ref{Tab3}, but for the observed
methanol lines in the Orion-KL spectrum.
}
\label{Tab4}
\begin{tabular}{c r@.l  c c r@.l r@.l}
\hline\\[-12pt]
\# & \multicolumn{2}{c}{$f_{\rm obs}$} & $FWHM$ & 
$T^{\scriptscriptstyle\rm peak}_{\scriptscriptstyle\rm A}$ &
 \multicolumn{2}{c}{$V^{\scriptscriptstyle (1)}_{\scriptscriptstyle\rm LSR}$} &
 \multicolumn{2}{c}{$V^{\scriptscriptstyle (2)}_{\scriptscriptstyle\rm LSR}$} \\
 & \multicolumn{2}{c}{(MHz)} & (\kms) & (K) & 
  \multicolumn{2}{c}{(\kms)} &
   \multicolumn{2}{c}{ (\kms)} \\
\hline\\[-10pt]

1&493685&675(7) A& 4 & 8.4(3) & 8&15(4) & 8&16(4) \\
2&493720&247(7) A& 4 & 8.6(3) & 8&15(4) & 8&16(4) \\
3&494468&182(8) A& 4  & 5.8(2) & 8&19(4) & 8&11(4) \\
4&495159&606(8) E& 4  & 6.0(2) & 8&17(4) & 8&17(4) \\[5pt]

5&541986&222(8) A& 4 & 9.7(4) & 8&16(4) & 8&15(4) \\
6&542067&209(8) A& 4 & 9.7(4) & 8&14(4) &  8&15(4) \\
7&543061&376(12) E& 4 & 6.7(3) & 8&18(4) & 8&17(4) \\[5pt]

8&590261&731(4) A& 4 & 10.5(4) & 8&10(4) & 8&12(4) \\
9&590424&408(4) A& 4 & 10.6(4) & 8&06(4) &  8&14(4) \\
10&590774&822(7) E& 5 & 7.4(3) & 8&19(4) & 8&18(4) \\[5pt]

11&638262&142(13) E& 5 & 7.5(2) & 8&18(3) & 8&22(3) \\
12&638506&307(8)  A&  4 & 11.1(3) & 8&07(3) & 8&08(3) \\
13&638800&633(8) A& 4 & 11.0(3) & 8&07(3) & 8&07(3) \\
\hline
\end{tabular}
\end{table*}

Before proceeding directly to the estimates of $\Delta\mu/\mu$,
which are obtained from various combinations of A- and E-methanol lines,
it is necessary to analyze the measured sequence of peak temperatures to determine the physical conditions under which
this sequence could form.
This question arises due to the fact that direct radiation or collision
processes are prohibited between A- and E-methanol, and the 
A/E abundance ratio  in the molecular
cloud practically does not change from the moment of formation of the molecules.
At the same time, the formation of methanol in a cold gas at 
$T_{\rm kin} = 10-15$~K
leads to an overabundance of A-methanol, but in a warm or hot gas at 
 $T_{\rm kin} \ga 40$~K the  
 A/E ratio is aligned \cite{Wir}.
Therefore, in some cases, these two forms of methanol are considered as
two molecules and it needs to be checked whether they trace the same gas.

\begin{figure}
\vspace{-4.0cm}
\centering
\includegraphics[width=0.8\textwidth]{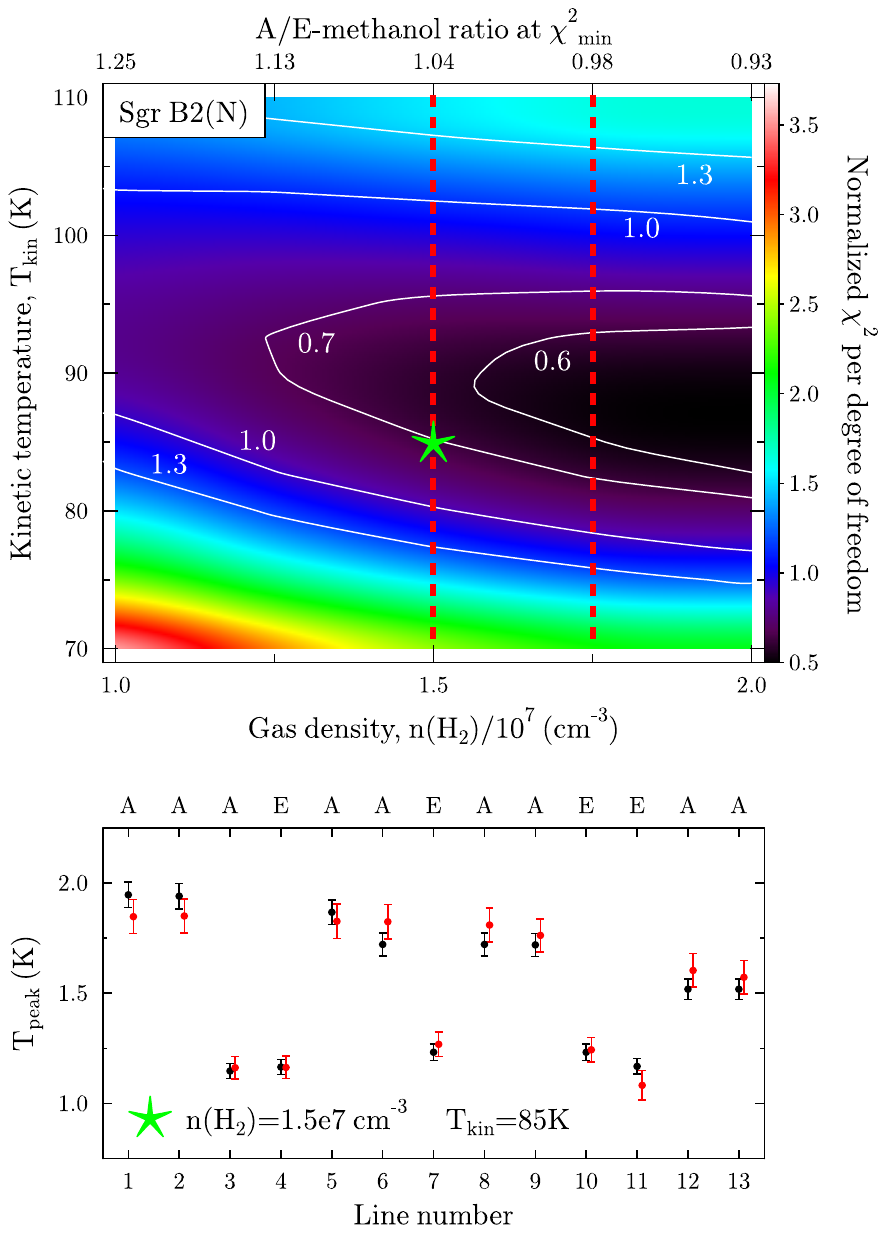}
\vspace{-1.5cm}
\caption{\small  
({\it Top panel.}) Map of normalized
minimum values of $\chi^2_{\rm min}$ as a function
of gas density $n(H_2)$ and kinetic temperature $T_{\rm kin}$
at different values of column densities $N_{\scriptscriptstyle\rm A}$ and
$N_{\scriptscriptstyle\rm E}$. Vertical dotted lines limit
the range of acceptable values of A/E.
The point is marked with a star 
\{$n(H_2), T_{\rm kin}, N_{\scriptscriptstyle\rm A}, N_{\scriptscriptstyle\rm E}$\} =
\{1.5e7, 85, 1.96e15, 1.86e15\}.\\
 ({\it Bottom panel.}) Example of calculated and observed peak temperatures (red and black, respectively) 
 for the thirteen methanol lines from
Table~\ref{Tab3} at the point marked with the star in the upper panel.
The minimum values of $\chi^2$ at this point are as follows:
$\chi^2_{\scriptscriptstyle\rm A} =0.7$, $\chi^2_{\scriptscriptstyle\rm E}=0.8$, and
$\chi^2_{\scriptscriptstyle\rm A+E} = 0.7$.
}
\label{Fig2}
\end{figure}

Peak temperatures from Table~\ref{Tab3} were checked 
 using the RADEX code \cite{Tak},
which calculates the radiation transfer in molecular lines when deviating from local thermodynamic equilibrium (non-LTE).
A model of a homogeneous plane-parallel layer was used, which most closely describes the observed intensities of the methanol lines.
We took the temperature of the background radiation as 
 $T_{\rm bg} = 2.73$~K, and as the line width~-- the median value
from Table~\ref{Tab3}, $FWHM = 14$ km~s$^{-1}$.
The collision rates  with H$_2$ are known with an error 
of 3\% \cite{RF},
which was taken into account when  
comparison of the calculated peak temperatures with the observed ones.
It turned out that the observed sequence of peak temperatures can be matched with model values only in a narrow range of gas densities $1.5\times10^7$ cm$^{-3}$ $\leq n(H_2)\leq 1.75\times10^7$ cm$^{-3}$  
 and kinetic temperatures of 80~K $\leq T_{\rm kin} \leq$100~K.
This area is marked by two dotted vertical lines in Fig.~\ref{Fig2}.
As the gas density decreases, $n(H_2) < 1.5\times10^7$ cm$^{-3}$,
the A/E ratio becomes greater than one, and when 
$n(H_2) > 1.75\times10^7$ cm$^{-3}$ this ratio decreases, A/E $<1$.
If one decreases or increases the kinetic temperature inside the selected area, then $\chi^2$ increases dramatically.

The lower panel in Fig.~\ref{Fig2} illustrates one of the possible solutions, the position of which is marked with a star on the top panel.
The numbers of the methanol lines are plotted along the lower horizontal axis according to Table~\ref{Tab2}, with the corresponding type of methanol at the top~-- A or E.
The measured peak temperatures and their uncertainty intervals are marked in black, which include systematic temperature scale calibration errors (4\% for bands 1a, 1b and 
3\% for bands 2a \cite{HIFI}) 
and the average noise amplitude ($\sigma_{\rm rms}$).
The red color corresponds to the calculated peak temperatures.

The relatively low values of $\chi^2$ in the range of acceptable values
of physical parameters reflect the fact that we used 
major estimates of temperature scale calibration errors.
However, this does not
affect the main conclusion that the selected A- and E-methanol lines trace the same
gas regions with similar physical parameters 
 in the molecular complex Sgr\,B2(N).

\section{Estimates of $\Delta\mu/\mu$}
\label{Sec4}

To perform differential measurements of the fundamental physical constant $\mu$, it is necessary to have pairs of methanol lines with different sensitivity coefficients $Q_i$, $Q_j$ and measured radial velocities 
$V^{(k)}_i$, $V^{(k)}_j$ \cite{LKR}:
 \begin{equation}
(\Delta\mu/\mu) \equiv (\mu_{\rm obs} - \mu_{\rm lab})/\mu_{\rm lab} =
(V^{\scriptscriptstyle (k)}_i - V^{\scriptscriptstyle (k)}_j)/[c(Q_j - Q_i)]\, ,
\label{E3000}
\end{equation}
where $c = 299792.458$ km~s$^{-1}$~ is the speed of light,
 $k$ = 1 or 2 if laboratory or calculated frequencies are used, respectively.

In order to avoid the various systematic errors mentioned in 
Sect.~\ref{Sec2}, we compared sets of lines that fall into the same subrange.
In addition, to avert the correlation in errors of
$\Delta\mu/\mu$ (see \cite{VLH, L10}),
we use the average values of radial velocities for lines with similar coefficients $Q$.
In Table~\ref{Tab5} this is marked on the top panel:
the designations $V_1$ and $V_2$ indicate the radial
velocities of the lines being compared, and the numbers under these designations indicate for which lines the average value was calculated  
(the numbering corresponds to Table~\ref{Tab1}).
For example, writing 1+2 means that the average velocity
was calculated from the radial velocities of lines 1 and 2.
The average values of the sensitivity coefficients of the compared lines $Q_1$ and $Q_2$ are shown below.
In Table~\ref{Tab5} the average velocities using both laboratory and calculated frequencies are given, and the 
corresponding calculated values of
$\Delta\mu/\mu$ in units of $10^{-7}$ are
 shown as well.
This table also shows weighted mean estimates of $\langle\Delta\mu/\mu\rangle^{(k)}$ obtained using laboratory ($k=1$) or calculated ($k=2$) reference frequencies.

The weighted mean estimates of $\langle\Delta\mu/\mu\rangle^{(k)}$ were calculated using the method of processing an
unequal accuracy data set  \cite{A72}.
For Sgr\,B2(N) in the central part of the Galaxy  
$\langle \Delta\mu/\mu \rangle^{(1)} = (-4.1 \pm 0.6) \times 10^{-7}$ 
($6.8\sigma$ C.L.)\footnote{C.L. means confidence level.},
if laboratory frequency measurements are used, or 
$\langle \Delta\mu/\mu \rangle^{(2)} = (-3.5 \pm 0.5) \times 10^{-7}$
($7\sigma$ C.L.),
if calculated frequencies are used instead of laboratory ones.
In any case, the result indicates a reduced value of $\mu$ with a high level
of statistical significance.
In this case, the $\kappa$ parameter indicating the presence
of unaccounted-for systematic errors in the sample
or misses at $\kappa > 2$ (equation (4.138) in
\cite{A72}), in our case equal to 1.17 and 0.98 for the first and second 
 the samples of \dmm, respectively.

At the same time, for Orion-KL~-- an object far from the Galactic Center, 
$\langle \Delta\mu/\mu \rangle^{(1)} = (-1.4  \pm 0.6) \times 10^{-7}$
($2.3\sigma$ C.L.) according to laboratory frequencies and 
$\langle \Delta\mu/\mu \rangle^{(2)} = (-1.1 \pm 0.8) \times 10^{-7}$
($1.4\sigma$ C.L.) if calculated frequencies are utilized.
The obtained results for Orion-KL show only the upper limit on the variation of $\mu$ at the level of 
$\Delta\mu/\mu < (6-8) \times 10^{-8}$.

Presented in Table~\ref{Tab5} the \dmm\ values
based on laboratory methanol frequencies
can be combined with similar data from Table~3 in \cite{VLH}, which
contains \dmm\ measurements from observations at the IRAM 30-m telescope of two regions of Sgr\,B2(N) and (M).
In this case, the sample size increases
to $n = 14$ and the following estimates are obtained:
$\langle \Delta\mu/\mu\rangle^{(1)} = (-3.4 \pm 0.4) \times 10^{-7}$
($8.5\sigma$ C.L.) and $\kappa = 0.13$.

A statistical normality test criterion can be applied to the combined sample of size $n = 14$.
For small size samples
$(n \sim 10)$ the Shapiro-Wilk criterion is considered to be the most effective \cite{ShW}\footnote{The Shapiro-Wilk statistic $W$ lies in
the range $0 < W < 1$, the right boundary of which indicates that the data is close to normal. 
At the same time, in order to have no reason to reject 
the hypothesis of normality,
a given level
of significance, for example, $\alpha = 0.05$ 
(confidence probability 95\%) must correspond to
$p$-value $> 0.05$.}.
In our case, $W = 0.94$ and $p$-value = 0.36 at $\alpha =0.05$,
i.e., the sample is close to normal.

Since all
the systematic effects known to us have been taken into account in the course of this study, the result obtained  in the direction of the Galactic center
$\langle \Delta\mu/\mu \rangle^{(1)} = (-3.4 \pm 0.4) \times 10^{-7}$ may indicate a possible decrease in the parameter $\mu$ compared to its laboratory value.
As confirmation of the absence of any systematics, 
the spectra of the Orion-KL molecular cloud located 
far from the Galactic Center, were analyzed and 
no reliable variations of $\mu$ have been found.

\begin{table*}[h!]
\centering
\caption{Estimates of $\Delta\mu/\mu$ (in units of $10^{-7}$)
in Sgr\,B2(N) and Orion-KL along the lines of methanol \chhhoh\
(see Eq.~(\ref{E3000})).
Measured average values of radial velocities
 $V_1$ and $V_2$ (in km~s$^{-1}$)
are grouped according to Table~\ref{Tab2} in four spectral
ranges.
The numbers of lines with approximately the same
sensitivity coefficients are located under the symbols $V_1$ and $V_2$, 
and the average
values of the sensitivity coefficients themselves are given in the row below.
The $1\sigma$ errors in the last digits are shown in parentheses.
}
\label{Tab5}
\begin{tabular}{ |c c |c c| c c | c c | }
\hline\\[-12pt]

\multicolumn{1}{|c}{$V_1$} & \multicolumn{1}{c|}{$V_2$} & 
\multicolumn{1}{|c}{$V_1$} & \multicolumn{1}{c|}{$V_2$} & 
\multicolumn{1}{|c}{$V_1$} & \multicolumn{1}{c|}{$V_2$} & 
\multicolumn{1}{|c}{$V_1$} & \multicolumn{1}{c|}{$V_2$} \\[-4pt]
\multicolumn{1}{|c}{$\scriptscriptstyle 1+2$} & 
\multicolumn{1}{c|}{$\scriptscriptstyle 3+4$} & 
\multicolumn{1}{|c}{$\scriptscriptstyle 5+6$} & 
\multicolumn{1}{c|}{$\scriptscriptstyle 7$} & 
\multicolumn{1}{|c}{$\scriptscriptstyle 8+9$} & 
\multicolumn{1}{c|}{$\scriptscriptstyle  10$} & 
\multicolumn{1}{|c}{$\scriptscriptstyle 11$} & 
\multicolumn{1}{c|}{$\scriptscriptstyle 12+13$} \\[-4pt]
\multicolumn{1}{|c}{$\scriptscriptstyle Q_1=-0.1$} & 
\multicolumn{1}{c|}{$\scriptscriptstyle Q_2=1.65$} &
\multicolumn{1}{|c}{$\scriptscriptstyle Q_1=0.0$}  &
 \multicolumn{1}{c|}{$\scriptscriptstyle Q_2=1.7$}  &
\multicolumn{1}{|c}{$\scriptscriptstyle Q_1=0.1$} & 
\multicolumn{1}{c|}{$\scriptscriptstyle Q_2=1.6$} &
\multicolumn{1}{|c}{$\scriptscriptstyle Q_1=1.6$}  & 
\multicolumn{1}{c|}{$\scriptscriptstyle Q_2=0.1$}  \\
\hline
\hline

\multicolumn{8}{|c|}{\it Sagittarius\,B2(N), $f_{obs}$ \& $f_{lab}$ results}\\
\hline
{\footnotesize 65.14(21)} & {\footnotesize 65.42(5)} & 
{\footnotesize 65.12(4)} & {\footnotesize 65.34(6)} & 
{\footnotesize 65.43(3)} & {\footnotesize 65.61(6)} &
{\footnotesize 65.32(10)} & {\footnotesize 65.15(6)} \\
\multicolumn{2}{|c|}{\footnotesize {$\Delta\mu/\mu=-5.4\pm4.2$}} &
 \multicolumn{2}{|c|}{\footnotesize {$ \Delta\mu/\mu=-4.2\pm1.4$}} &
\multicolumn{2}{|c|}{\footnotesize {$\Delta\mu/\mu=-3.9\pm1.5$}} &
 \multicolumn{2}{|c|}{\footnotesize {$\Delta\mu/\mu=-3.7\pm2.6$}} \\
\hline\\[-12pt]
\multicolumn{8}{|c|}{{\it weighted mean} 
$\langle \Delta\mu/\mu \rangle^{\scriptscriptstyle (1)} = -4.1\pm0.6$ }\\
\hline
\hline

\multicolumn{8}{|c|}{\it Sagittarius\,B2(N), $f_{obs}$ \& $f_{cal}$ results}\\
\hline
{\footnotesize 65.15(21)} & {\footnotesize 65.38(4)} & 
{\footnotesize 65.12(3)} & {\footnotesize 65.32(5)} & 
{\footnotesize 65.47(3)} & {\footnotesize 65.60(5)} &
{\footnotesize 65.36(10)} & {\footnotesize 65.16(6)} \\
\multicolumn{2}{|c|}{\footnotesize {$ \Delta\mu/\mu=-4.4\pm4.2$}} & 
\multicolumn{2}{|c|}{\footnotesize {$\Delta\mu/\mu=-3.9\pm1.1$}} &
\multicolumn{2}{|c|}{\footnotesize {$ \Delta\mu/\mu=-2.9\pm1.1$}} & 
\multicolumn{2}{|c|}{\footnotesize {$\Delta\mu/\mu=-4.4\pm2.6$}} \\
\hline\\[-12pt]
\multicolumn{8}{|c|}{{\it weighted mean} 
$\langle \Delta\mu/\mu \rangle^{\scriptscriptstyle (2)} = -3.5\pm0.5$ }\\
\hline\\[-7pt]
\hline

\multicolumn{8}{|c|}{\it Orion-KL, $f_{obs}$ \& $f_{lab}$ results}\\
\hline
{\footnotesize 8.15(3)} & {\footnotesize 8.18(3)} & 
{\footnotesize 8.15(3)} & {\footnotesize 8.18(3)} & 
{\footnotesize 8.08(3)} & {\footnotesize 8.19(4)} &
{\footnotesize 8.18(4)} & {\footnotesize 8.07(2)} \\
\multicolumn{2}{|c|}{\footnotesize {$ \Delta\mu/\mu=-0.5\pm0.8$}} & 
\multicolumn{2}{|c|}{\footnotesize {$ \Delta\mu/\mu=-0.5\pm1.0$}} &
\multicolumn{2}{|c|}{\footnotesize {$\Delta\mu/\mu=-2.3\pm1.0$}} & 
\multicolumn{2}{|c|}{\footnotesize {$\Delta\mu/\mu=-2.6\pm0.9$}} \\
\hline\\[-12pt]
\multicolumn{8}{|c|}{{\it weighted mean} 
$\langle \Delta\mu/\mu \rangle^{\scriptscriptstyle (1)} = -1.4\pm0.6$ }\\
\hline
\hline

\multicolumn{8}{|c|}{\it Orion-KL, $f_{obs}$ \& $f_{cal}$ results}\\
\hline
{\footnotesize 8.16(3)} & {\footnotesize 8.14(3)} & 
{\footnotesize 8.15(3)} & {\footnotesize 8.17(4)} & 
{\footnotesize 8.13(3)} & {\footnotesize 8.18(4)} &
{\footnotesize 8.22(3)} & {\footnotesize 8.07(2)} \\
\multicolumn{2}{|c|}{\footnotesize {$ \Delta\mu/\mu=0.4\pm0.8$}} & 
\multicolumn{2}{|c|}{\footnotesize {$\Delta\mu/\mu=-0.4\pm1.0$}} &
\multicolumn{2}{|c|}{\footnotesize {$\Delta\mu/\mu=-1.2\pm1.0$}} & 
\multicolumn{2}{|c|}{\footnotesize {$\Delta\mu/\mu=-3.4\pm0.9$}} \\
\hline\\[-12pt]
\multicolumn{8}{|c|}{{\it weighted mean} 
$\langle \Delta\mu/\mu \rangle^{\scriptscriptstyle (2)} = -1.1\pm0.8$ }\\
\hline
\end{tabular}
\end{table*}

\section{Discussion of the results and conclusion}
\label{Sec5}

\begin{figure}
\vspace{-10.0cm}
\centering
\includegraphics[width=0.8\textwidth]{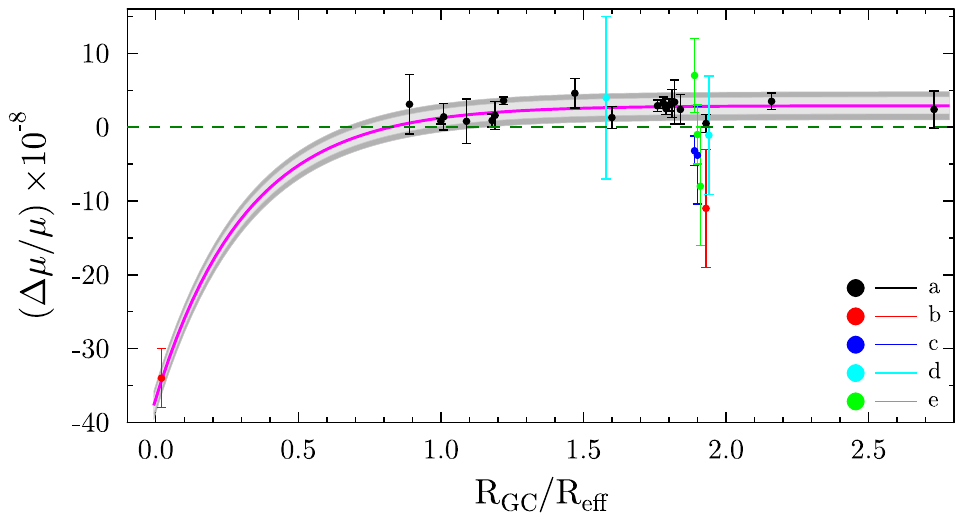}
\vspace{-0.5cm}
\caption{\small Measured values of \dmm\ along the galactocentric
radius $R_{\scriptscriptstyle\rm GC}$, normalized to the effective
Galactic radius $R_{\scriptscriptstyle\rm eff} \approx 4.5$ kpc.
The points with uncertainty intervals $\pm1\sigma$ are taken from the following publications:
a~--\cite{L22}, b~-- this work, c~--\cite{Dap}, d~--\cite{VKL24, VL24} and
e~--\cite{VLK26}, in which the values of \dmm\ were estimated 
along the lines
\chhhoh~(the first three links), $^{13}$CH$_3$OH 
(the next two links) and CH$_3$CHO (the last link).
The regression line (pink) is surrounded by the confidence
zones $\pm1\sigma$ (light gray) and $\pm3\sigma$ (dark gray).
}
\label{Fig3}
\end{figure}

Summing up the work done, it is interesting to compare  
the \dmm\ values in the Sgr\,B2(N), and Orion-KL spectra compare with those obtained earlier for
various molecular clouds distributed
in the Galactic disk in the range from $R_{\scriptscriptstyle\rm GC}\approx
4$ kpc to $R_{\scriptscriptstyle\rm GC} \approx 12$ kpc.
This comparison is illustrated in Fig.~\ref{Fig3}, where the points with
uncertainty intervals $\pm1\sigma$ show estimates of \dmm,
which used various lines of complex organic molecules CH$_3$OH,
 $^{13}$CH$_3$OH and CH$_3$CHO. These lines were observed on various
ground-based telescopes and at the {\it Herschel} space observatory.
Two statistical significance zones $\pm1\sigma$ and $\pm3\sigma$ are shaded in light gray and dark gray, respectively, for the regression
curve shown in pink.
The slight excess of this curve over the zero value of \dmm\
depends on the choice of reference frequencies (see Table~4 in \cite{L22}) and is not statistically significant.
Therefore, the main purpose of such a regression is simply
to graphically reflect a sharp change in the value of \dmm\ 
within the interval
$0 < R_{\scriptscriptstyle\rm GC}/R_{\rm eff} \la 1$.

A similar sharp increase in the circular velocities of stars
in the same range of galactocentric distances is observed in our
own and other spiral galaxies (e.g., \cite{Isi24, Isi25, Ei19}).
At the same time, it is generally assumed that in the range of 
$0 <R_{\scriptscriptstyle\rm GC}/R_{\rm eff} \la 1$
there is an increase in the mean density of dark matter against the background of a decrease in the mean
density of ordinary baryonic matter, 
and the alignment of $\rho_{\rm br}$ and
$\rho_{\scriptscriptstyle\rm DM}$ occurs when 
$R_{\scriptscriptstyle\rm GC}/R_{\rm eff} \ga 1$, 
after which, as $R_{\scriptscriptstyle\rm GC}$ increases, 
dark matter dominates, forming a Galactic halo with a virial mass 
$M_{\rm vir} \approx 7\times10^{11}M_\odot$ \cite{Ei19}.
In the region of $R_{\scriptscriptstyle\rm GC}/R_{\rm eff}\ga 1$, 
flat stellar rotation curves and zero
values of \dmm\ are observed at the level of a few $\times 10^{-8}$.

In addition, we note that the results of this work limit a
number of theoretical models proposed to explain possible
spatial changes in fundamental physical constants.
For instance, in theories considering additional
scalar fields that trace inhomogeneities
of the gravitational field,
the change in $\mu$ is associated with the gradient 
of the gravitational potential,
$\Delta\mu/\mu = k_\mu \Delta \Phi$ (e.g., \cite{MBS, FS, Blat}).
This relationship is not confirmed by measurements \dmm\  in Sgr\,B2(N) and Orion-KL,
since both molecular complexes have comparable gravitational potentials, 
$\Phi = GM/c^2r \sim 10^{-9}$ (here $r$ is the radius of the object),
but  they differ significantly in
the estimates of \dmm.

Other models (e.g., \cite{Kh, OP, Brax}),
which are based on the idea of shielding the fifth force
depending on the local density of baryonic matter, which differs by
many orders of magnitude between the interstellar medium and its values in terrestrial laboratories, are also not supported by these measurements of \dmm\ in Sgr\,B2(N) and Orion-KL.
The local densities of baryonic matter in these
clouds are comparable, $n(H_2) \sim 10^7$ cm$^{-3}$, 
but the estimates of \dmm\ are different.

So, since the \dmm\ estimates turn out, as we have shown, to be insensitive
neither to the values of the gravitational potential nor to the local density
of baryonic matter,
the conclusion suggests itself that the detected possible
correlation in the distributions of
$\rho_{\scriptscriptstyle\rm DM}/\rho_{\rm br}$ and \dmm\  
along the Galactic disk may be
related  to the density of dark matter.
Moreover, if dark matter modulates the Higgs scalar field,
then spatial changes in $\mu$  can be expected.
A similar scenario has been considered, for example, in 
multidimensional Kaluza-Klein models in \cite{Car21}.

\bigskip\noindent
In conclusion, we list the main results obtained in this work.

(1) Processing of two identical sets of high-frequency
methanol lines (490-640 GHz) in the spectra of two molecular clouds
Sgr\,B2(N) and Orion-KL
show the systematic shifts of these lines relative
to each other 
in the first case, and the absence in the second.

(2) Interpretation of this behavior of line frequencies in terms
of small  $\mu$-variations, 
taking into account all systematic corrections, 
leads to the following mean values:
$\langle\Delta\mu/\mu\rangle = (-3.4\pm0.4)\times10^{-7}$ for Sgr\,B2(N) and (M), and
$\langle\Delta\mu/\mu\rangle = (-1.1\pm0.8)\times10^{-7}$ for Orion-KL.

(3) It is shown using the RADEX code that the A- and E-methanol lines in Sgr\,B2(N) trace the same gas with the parameters
$n(H_2)\approx 1.5\times10^7$ cm$^{-3}$, $T_{\rm kin} \approx 85$~K,
$N_{\scriptscriptstyle\rm A}\approx 2\times10^{15}$ cm$^{-2}$ and the  abundance ratio A/E~$ \approx 1$ expected for hot gas.

(4) A comparison of the distribution of \dmm\ values along the galactocentric
radius with flat rotation curves of stars in the Galaxy
 demonstrates a similar nature of these two distributions with a sharp increase in the density of dark matter within the range 
$0 <R_{\scriptscriptstyle\rm GC}/R_{\rm eff} \la 1$ and the same
sharp change in the value of \dmm\ in the same range.
Such a correlation may indicate dark matter modulation of the Higgs scalar field.

\subsection*{ACKNOLEDGMENTS}
The authors would like to thank both anonymous
referees for their constructive comments.

\subsection*{FUNDING}
This work was supported by the Ministry of Science and Higher
Education of the Russian Federation, state assignment
FFUG-2024-0002 for the Ioffe Institute.

\subsection*{CONFLICT OF INTEREST}
The authors of this work declare that they have no conflicts of interest.

\end{document}